\documentclass[twocolumn,showpacs,amssymb,aps,nofootinbib,floatfix]{revtex4-1}
\usepackage{epsfig}
\newcommand{\be}{\begin{eqnarray}}
\newcommand{\ee}{\end{eqnarray}}
\usepackage{graphicx}

\newcommand{\order}[1]{ \mathcal{O} \left( #1 \right) }

  \newcommand{\lqcd}{\Lambda_{QCD}}

\newcommand{\dde}{\mathrm{d}\hspace{0.1em}}

\begin{document} \hbadness=10000
\topmargin -0.8cm\oddsidemargin = -0.7cm\evensidemargin = -0.7cm
\title{A percolation transition in Yang-Mills matter at finite number of colours}
\author{Stefano Lottini$^a$, Giorgio Torrieri$^b$}
\affiliation{$\phantom{A}^a$ ITP and $\phantom{A}^b$ FIAS,  J.~W.~Goethe Universit\"at, Max-von-Laue-Str.~1, 60438 Frankfurt am Main, Germany. }
\date{\today}

\begin{abstract}
We examine baryonic matter at quark chemical potential of the order of the confinement scale, $\mu_q\sim \lqcd$.
In this regime, quarks are supposed to be confined but baryons are close to the ``tightly packed limit'' where they nearly overlap in configuration space.
We show that this system will exhibit a percolation phase transition {\em when varied in the number of colours} $N_c$: at high $N_c$, large distance correlations at quark level are possible even if the quarks are essentially confined.
At low $N_c$, this does not happen.
We discuss the relevance of this for dense nuclear matter, and argue that our results suggest a new ``phase transition'', varying $N_c$ at constant $\mu_q$.
\end{abstract}
\maketitle

Long ago, it was suggested to describe the deconfinement phase transition via percolation \cite{perc1,perc2,perc3,perc4,perc5,perc6,perc7}.
The idea is that, at increasing energies, the high parton density will make partons of different hadrons overlap.
It is natural to associate this transition to  deconfinement, where a quark can propagate throughout the hot medium rather than being confined to the hadron size $\sim \lqcd^{-1} \sim 1\,\mathrm{fm}$ in natural units \cite{florkowski}. Currently, we believe that the transition to quark-gluon plasma at low density is a cross-over, casting doubt on the relevance of percolation (a phase transition, generally of second order)  in those conditions.

In this work we use the percolation picture to study a different but related region in the phase diagram, the one at low to moderate temperature $0 \le T \leq \lqcd$ and high quark chemical potential $\mu_q \sim \lqcd$.
Strongly interacting matter in this regime has recently received a considerable amount of interest.
Such matter can hopefully be produced in heavy-ion collisions \cite{low1,low2,low3,low4}, and is thought to exhibit a rich phenomenology, such as a critical point, spinodal instabilities \cite{spino}, separation between chiral symmetry and confinement \cite{hansen,ratti,dirkq,sasaki}, chirally inhomogeneous phases \cite{kojo,chiralspiral,buballa}, new phases \cite{quarkyonic}, etc.

\begin{figure}[t]
\epsfig{width=6.9cm,figure=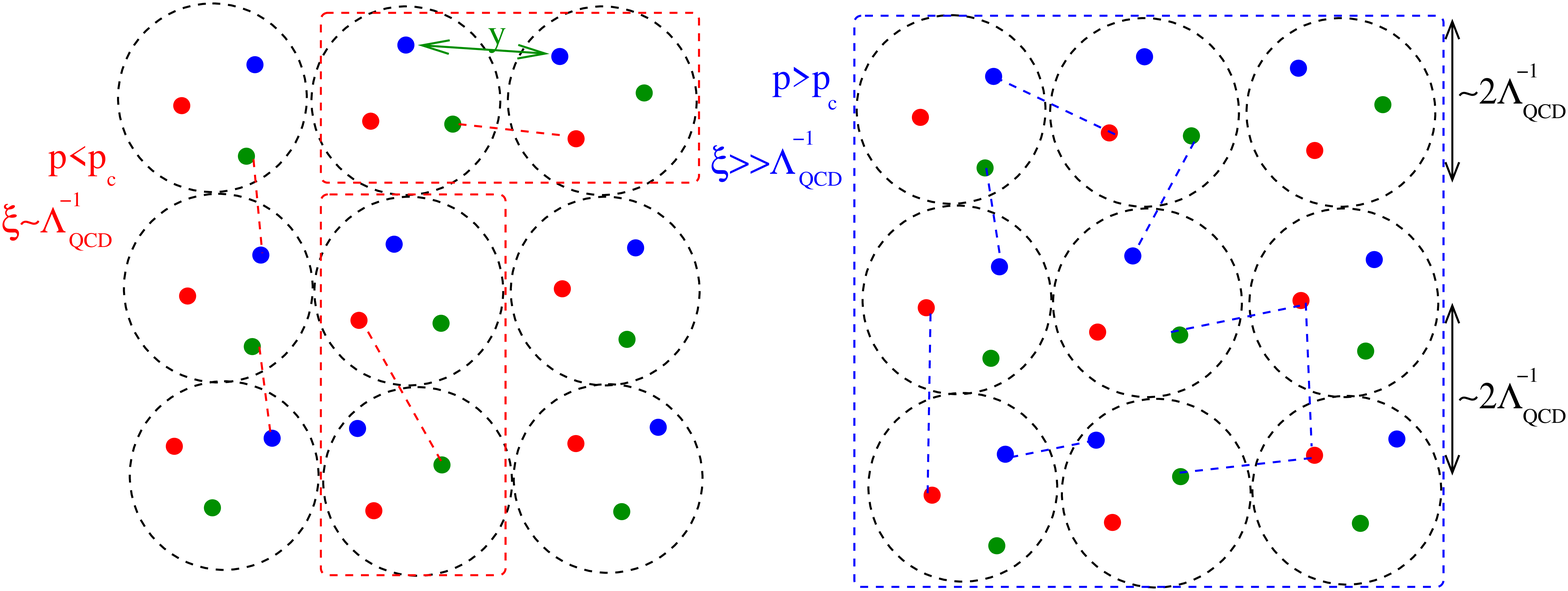}
\caption{\label{figpacking} (colour online) 
The structure, in configuration space, of dense baryonic matter. Comparison of the top and the bottom panels shows that the percolation picture is applicable: when the exchange probability goes above the threshold $p_c$ the size of the typical cluster diverges. The figure also shows a definition of the inter-quark distance $y$ of Eq.~\ref{propagator}.}
\end{figure}

These conjectures are, however, extraordinarily difficult to explore quantitatively in a rigorous manner.
The quark chemical potential $\mu_q$ is nowhere near the asymptotic freedom limit where perturbative QCD can be used \cite{florkowski}.
It is, however, way too high for existing lattice-based approaches, dependent on $\mu_q/T \ll 1$, to work \cite{latmu1,latmu2,latmu3}.
Hence, a simple geometric picture such as percolation might help.
Its physical relevance is demonstrated in Fig.~\ref{figpacking}, schematically showing the regime where $\mu_q \sim \lqcd$.
At this chemical potential the density is naively expected to be $\sim \order{1}\lqcd^3$, that is, one baryon per baryonic size.
In configuration space, this means baryons touch each other, i.e.~their quarks are separated by a scale {\em not} much larger than the confinement scale.
We therefore expect some weakly-coupled features of QCD be present due to the small inter-quark distance.
Non-perturbative features, on the other hand, should also be present since baryons are still confined.
The interplay of all these features could be very physically interesting.
Such a setup was recently investigated in \cite{quarkyonic} using the only relevant quantity that can be called ``a small parameter'': $1/N_c$, with $N_c$ number of colours \cite{thooft}.
The idea is to keep the quantity $\lambda=g^2 N_c$ (where $g$ is the $SU(N)$ Yang-Mills coupling constant \cite{florkowski}) fixed, sending the number of colours $N_c$ to infinity, and then expanding in $1/N_c$.
It is easy to see that the running of $\lambda$ is qualitatively $N_c$-independent, hence $\lqcd \sim N_c^0$.
Obviously, such a set-up can give at best a qualitative agreement at $N_c=3$, but it might be enough to understand the phase diagram structure of the system.
In this setup each quark-quark interaction is {\em weak} ($\sim \lambda/N_c$) but, due to combinatorics, baryons remain strongly-coupled ``semiclassical'' objects \cite{witten}.
Looking at Fig.~\ref{figpacking} with a large-$N_c$ perspective, one can immediately see that quarks may be arbitrarily close together in configuration space (inter-quark distance $\sim N_c^{-1/3}$), so interactions of quarks between neighbouring baryons could be weakly coupled.
Confinement, however, persists until  quark-hole screening $\sim$ gluon antiscreening \cite{quarkyonic}, $\mu_q \sim N_c^{1/2} \lqcd$.

How can a confined system be, at the same time, asymptotically free?
The authors of \cite{quarkyonic} conjectured that excitations below the Fermi surface behave as asymptotically free quarks, but excitations close to the Fermi surface are confined.
Thus, the entropy of such system is $\sim N_c$ (unlike $N_c^0$ for confined matter), but dynamical excitations are baryons and mesons and the Polyakov loop expectation value \cite{florkowski} is $\sim 0$.
This new ``quarkyonic'' state of matter \cite{quarkyonic} might be realised in our $N_c=3$ world. 

The problem is that it is difficult to model this phase theoretically in a rigorous way.
While quarkyonic matter has been claimed to be found in the pNJL model \cite{sasaki}, one can not reliably investigate whether this phase has pressure $\sim N_c$ in that model, since the pressure in pNJL is {\em always} somewhat $\sim N_c$, since the ``Polyakov loop field'' as implemented in \cite{hansen,ratti,sasaki} is different from ``true'' confinement.
Similarly, it is difficult to see how holographic methods \cite{maldacena} can verify such a conjecture, since there $\lambda N_c^{-1}$ plays the role of the string coupling constant $g_s$ ($\lambda$ is the compactification scale of the remaining 5 dimensions in units of Planck length); in a {\em semiclassical} gravity regime, such as \cite{sugimoto}, pressure in a confined phase is always $\sim N_c^0$ \cite{lippert,rozali}.
Taking these difficulties into account, it is not clear how applicable is the large $N_c$ limit to our $N_c=3$ world.
Nuclear matter seems to look very different in that world compared to ours, as it is a tightly bound crystal of baryons \cite{witten,crystal1,crystal3} (unless what we call large-$N_c$ matter is actually the quarkyonic phase).
Mean field analysis show a considerable $N_c$ variation between $N_c=3$ and $10>N_c>3$ \cite{ncgiac}.
In \cite{ncus} a physical interpretation of these structures was conjectured: due to the Pauli exclusion principle, an interplay exists between the number of colours $N_c$ and the number of neighbours $N_N$.
Hence, a ``phase transition'' exists in $N_c$ space when the baryon density is $\sim \lqcd^3$, separating our world ($N_c \ll N_N$) from the truly large-$N_c$ world.
In this work, we aim to apply standard percolation knowledge to investigate this further.
\begin{figure*}[t]
\epsfig{height=5cm,figure=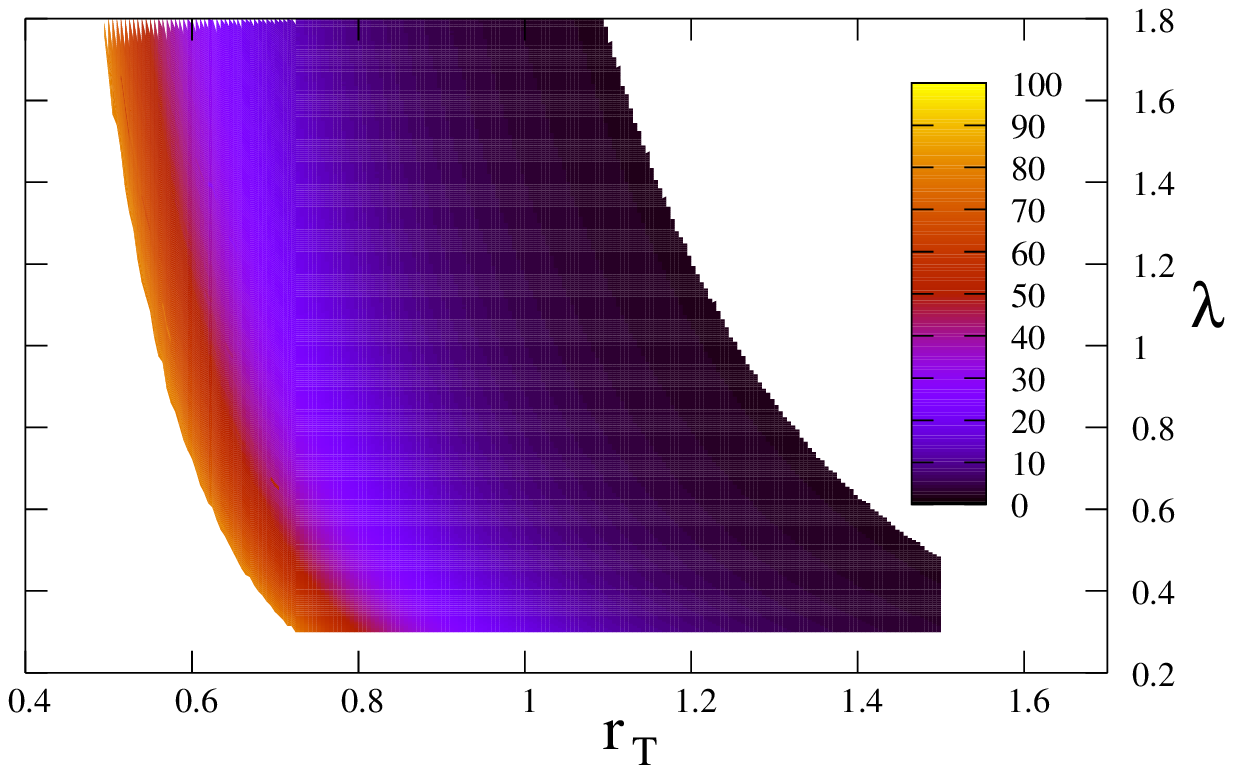}
\epsfig{height=5cm,figure=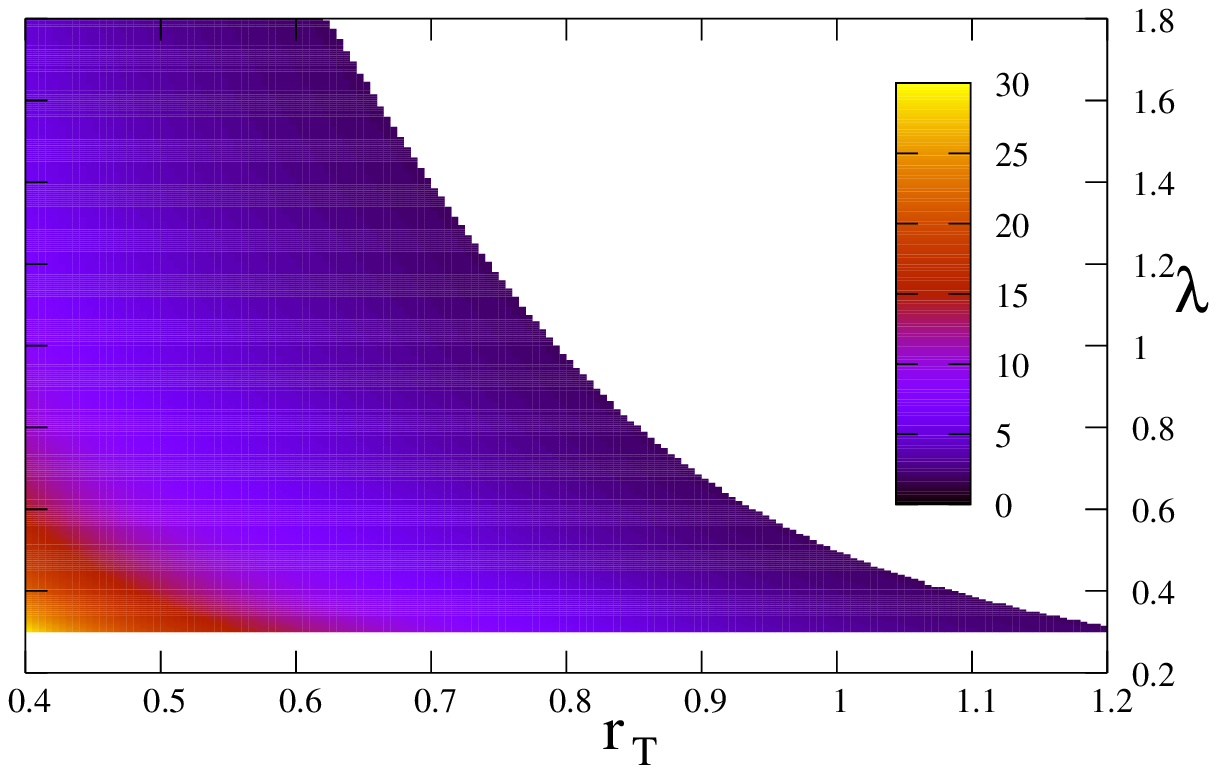}
\caption{\label{figperco} (colour online) Contour plot of the critical $N_c$ for the percolation transition in a hexagonally packed lattice  as a function of the coupling $\lambda$ and range $r_T$ (in $\lqcd$ units). The left panel assumes a $\Theta$-function correlation probability in configuration space, the right panel assumes a correlation probability based on the propagator used in \cite{kojo}.  Diagram covers $2 \le N_c \le 80$
}
\end{figure*}

The key insight suggesting that interesting structures might be lurking in $N_c$ is that 3D bond-percolation exhibits a phase transition at comparatively low critical link probability: for instance, $p_c\sim 0.25, 0.18, 0.12$ for simple-cubic (schematically shown in Fig.~\ref{figpacking}), body-centered-cubic and hexagonal-close-packed lattice, respectively \cite{percthresholds}.
Such values suggest that long-distance correlations on the quark level could occur even with a somewhat low percentage of quarks hopping between baryons, i.e.~firmly in the confined phase.
While below $p_c$ the characteristic correlation distance $\xi$ ($\sim$ cluster size) is $\sim \lqcd^{-1}$, above the threshold this quantity explodes to the total system size in a comparable amount of time. We leave the meaning of ``correlation''  vague, as it can be either a quark hop or a gluon exchange; in our context, it implies exchange of colour degrees of freedom within a confined tightly packed medium.
We encode the likelihood of exchange between neighbouring baryons in a link probability $p$, to be compared with the percolation threshold $p_c$ in order to assess the formation of large-scale structures.

Two baryons will be correlated if {\em at least two} quarks are correlated.
One has to sum over all possible multi-quark configurations, resulting in a strong $N_c$-dependence of $p$.
We determine the latter by calculating the probability $q=1-p$ that {\em no} exchanges happen between neighbouring baryons.
We define $p_{(1),ij}$ as the probability that quarks $i$ and $j$, respectively in baryons $A$ and $B$, ``correlate'' (either by flip or gluon exchange).
Assuming the quarks inside the nucleon are uncorrelated (Fermi motion dominates), this probability factorises into a geometric distribution $f_{A,B}(\mathbf{x})$ for quarks to be at a certain (vector) position $\mathbf{x}$ (see Fig.~\ref{figpacking}), and a ``squared propagator'' transition amplitude $F(d)$ for them:
\begin{eqnarray}
		\label{ptotlink}
	p &=& 1- \big(q_{(1),ij}\big)^{(N_c)^\alpha}\,;\\
	q_{(1),ij} &=& 
		\int  f_A (\mathbf{x}_i) \dde \mathbf{x}_i \int f_B(\mathbf{x}_j) \dde \mathbf{x}_j
		\left( 1- F(|\mathbf{x}_i-\mathbf{x}_j|)  \right)  \;. \nonumber
\end{eqnarray}
We assume a ``hard-sphere'' distribution for $f_{A,B}$
(since we keep $\mu_q$ fixed, the distance between centres of neighbouring baryons is always $2\lqcd^{-1}$, Fig.~\ref{figpacking}):
\begin{equation}
f_{A,B} (\mathbf{x}) \;\; \propto \;\;  \Theta \left( 1  - \lqcd \left| \mathbf{x} - \mathbf{x}_{A,B}^{\mathrm{centre}} \right| \right)
\end{equation}
 and a probability of exchange $i\leftrightarrow j$ based on a range of ``reasonable'' propagators, compatible with confinement (fast fall-out in configuration space at distances greater than $r_T \sim 1$ in units of $\lqcd^{-1}$) and with the large $N_c$ limit of QCD, the interaction is 
$\sim g^2 \sim \lambda/N_c$ \cite{thooft,witten}.
The propagators we use are the simple $\Theta$-function in configuration space and the momentum-space $\Theta$-function used in \cite{kojo}, all normalised 
so their area is $\lambda r_T/N_c$.  In configuration space the transition amplitudes are, respectively,
\begin{equation}
\label{propagator}
F(y) = \frac{\lambda}{N_c}  \left\{ 
\begin{array}{c}
\Theta(1 - \frac{y\lqcd}{r_T})\\
 \frac{2 r_T^2 }{\pi y^2} \sin^2\Big(\frac{y\lqcd}{r_T}\Big)
\end{array}
 \right.
\end{equation}
Other transition amplitudes, such as a Gaussian distribution in configuration space, were also tried with no significant modifications of the results presented below.

We note that, due to the fact that 3D percolation has a second-order phase transition at a certain $p_c < 1$, the results we obtain below have some 
degree of universality: as long as the qualitative features of confinement are observed (the transition amplitude $F(y)$ drops sharply above the scale $r_T$, and the hadron density profile $f_{A,B}(\mathbf{x})$ has a central plateau of radius $\sim \lqcd^{-1}$ and a sharply decreasing
tail outside), the results we show vary quantitatively but not qualitatively, in particular in regard to phase transition behavior if $p \sim p_c$
for some critical $N_c$. Mathematically, this model is similar to the Glauber model, familiar in heavy-ion collisions \cite{florkowski}, with
the number of colours playing the r\^ole of the number of participants $N_{part}$.
Just as in the Glauber model the dependence on $N_{part}$ is universal with respect to  cross-sections, in this case the dependence on $N_c$ is universal with respect to the shape of the transition amplitude.

The crucial parameter left is $\alpha$ in Eq.~\ref{ptotlink}.
Here, we shall consider two limits: one can assume that the link of quark $i$ from baryon A to quark $j$ of baryon B does not prejudice in any way the possibility of {\em also} linking $i$--$k$, with quark $k$ again in baryon B.
This scenario, natural if the link is actually realised by a gluon exchange rather than a quark flip, means that $\alpha = 2$.
If, on the other hand, the link is given by a quark exchange, then, on a short enough time-step, the probability of a quark moving more than once is negligible.
In this case, $\alpha = 1$.

Physically one expects that at $N_c \gg 1$ gluon exchange dominates over quark flip, by combinatorics alone.   Indeed, 
one can easily see that $\alpha = 1$ is in contradiction with the Skyrme crystal picture at large $N_c$:  in this picture, $p(N_c)$ approaches a constant large-$N_c$ value {\em from above}: {\em low $N_c$} nuclear matter would be more correlated (and hence more strongly bound) than {\em high $N_c$} nuclear matter.   Comparing strongly coupled $N_c \rightarrow \infty$ nuclear matter \cite{witten} to the weakly bound nuclear liquid at $N_c=3$ \cite{ncus}, this is obviously not right.
We therefore assume $\alpha=2$ henceforward.
A closer inspection of the $\alpha=2$ case reveals that it is an approximation of
\begin{eqnarray}
        p &=& 1 - \int  \prod_{\ell=1}^{N_c} f_A ( \mathbf{x}^{(A)}_\ell ) \dde \mathbf{x}^{(A)}_\ell
			\int f_B ( \mathbf{x}_{m}^{(B)} ) \prod_{m1}^{N_c} \dde \mathbf{x}^{(B)}_m \times \nonumber \\
		& & \times \prod_{i, j} \big( 1 - F(|\mathbf{x}^{(A)}_i-\mathbf{x}^{(B)}_j|)\big)\;,
\end{eqnarray}
where we  integrate over all quark positions in the two baryons and write the probability of no exchanges taking place as the product of the individual no-exchanges probabilities for $N_c^2$ $A$-$B$ pairs.
Numerical integration shows the effect of correlations to be a $\sim 3\%$ correction, so that the qualitative outcome of the analysis is unaffected.

In the large $N_c$ limit for the case $\alpha=2$, $p$ asymptotically approaches unity.
It is reasonable that this is the point where the ``dense baryonic matter as a Skyrme crystal'', theorised in \cite{witten,quarkyonic,crystal1,crystal3}, is reached.
If this is the case, however, one should remember that a percolation second-order phase transition occurs at a $p_c\ll 1$. 
Hence, keeping $\mu_q \sim \lqcd$ fixed but varying $N_c$, the features of the Skyrme crystal should manifest not with a continuous approach, rather as a second order transition at a not too high $N_c$, whose order parameter can be though to be the ``giant cluster'' density.
Below the critical $N_c$ there is little correlation between quarks of different baryons, while above this threshold they can correlate, with the distance boundary given only by causality.
We reiterate that this is {\em not} deconfinement since $\mu \sim \lqcd \ll N_c^{1/2} \lqcd$ independently of the number of colours, and the {\em fraction} of correlated quarks from different hadrons is still $\sim 0.1-0.3 \ll 1$ at the percolation transition. Right above this transition, therefore, the baryonic wavefunction should {\em not} be too different from the large-$N_c$ baryonic wavefunction described in \cite{witten}. The {\em correlation distance} of quarks will however be much larger than the baryon size.
The features of this new phase are therefore similar to those of the quarkyonic matter \cite{quarkyonic}.

Assuming the lowest 3D value $p_c=0.12$, appropriate for a closely packed hexagonal lattice, the critical number of colours is shown in Fig.~\ref{figperco} as a function of $\lambda$ and $r_T$.  As can be seen, the critical number of colours is significantly larger than three for $r_T \sim \lqcd^{-1},\lambda \sim 1$.   However, given the roughness of our model, a critical $N_c\leq 3$ can not be excluded at $\mu_q \sim \lqcd$, provided quarks can correlate significantly above the confinement length ($\sim 1.5 \lqcd^{-1}$) or the coupling constant is significantly larger than unity
(one has to remember, however, that these quantities are {\em not} independent: $\lqcd$ is defined as the scale at which $\lambda$ becomes ``strong'', $\sim 1$, and it is generally assumed that confinement is set by that scale).  

Considering Fig.~\ref{figperco} is a {\em lower limit} since $p_c$ is at its minimum ($p_c$ is significantly higher both in a Skyrme cubic crystal and in a disordered fluid), we can  say that $N_c=3$ is disfavoured, although it can not be excluded.
Changing temperature and $\mu_q$ should further change the critical $N_c$.   Exploring this parameter space, and seeing how it relates to the confinement phase transition, is the subject of a forthcoming work.

What are the phenomenological consequences of percolation?
If by ``correlation'' we mean energy-momentum-exchange via quark tunneling between baryons, it is reasonable that pressure and entropy density $\sim N_c$ above the percolation  threshold, while below it they stay $\sim N_c^0$.  This is because above the threshold, where interbaryon tunnelling is significant, "typical" excitations of the Fermi surface will be superpositions across baryons of baryon-localized quark-hole excitations, similar of conduction band electrons in a metal; while the localized excitation energy $\geq \lqcd$, the superposition makes its energy $\sim \lqcd$ even if color degeneracy remains. 
 Thus, the degrees of freedom of the system above percolation will be delocalized weakly interacting quarks in a lattice of confining potentials, a picture compatible with \cite{chiralspiral,quarkyonic}.
Below the threshold, where tunnelling is negligible, excitations are either color singlets or of energy $E \gg \lqcd$, suppressed below deconfinement.

The only known rigorous way to access these phenomena quantitatively is the lattice.
Current quenched simulations show that $N_c$ dependence is surprisingly smooth \cite{lat1,lat2}.
Our considerations suggest that this will not be true at finite chemical potential.
While chemical potentials accessible to current numerical studies are far smaller than dense-packing densities \cite{latmu1,latmu2,latmu3},  approaches such as the strong coupling expansion \cite{fromm} could be used to probe the large-$N_c$ dependence at these densities.   

This transition might also be visible with holographic techniques {\em beyond} the supergravity limit, since finite $\lambda N_c^{-1}$ corresponds, in gauge/string duality, to $g_s$ \cite{maldacena}.
Thus, percolation will manifest itself as a transition from  ``stringy weak coupling'' to ``not-so-weak-coupling''.   While there are hints \cite{stringperc,nicolini} that percolation is relevant for corrections beyond classical gravity, exploring this is beyond the scope of this work.  The fact that percolation appears only as a subleading factor of $g_s$ might explain why, despite the reasonableness of the argument in \cite{quarkyonic} for $s \sim N_c$ in the quarkyonic phase, $s \sim N_c^0$ in the confined phase in all semiclassical AdS/CFT setups to date.

In conclusion, we have used a toy model, with  universal features, to investigate the close-packed regime ($\mu_q \sim \lqcd$) of baryons at variable $N_c$.
Our findings suggest that, if baryons are kept in this regime but $N_c$ is varied, a percolation-type phase transition occurs at some critical $N_c \sim \order{10}$, probably but not certainly higher than 3.
This transition is quite distinct from deconfinement as the percentage of quarks propagating on superbaryonic distances is quite low.
Nevertheless, the typical correlation length will be much larger than $\lqcd^{-1}$.
Further work needs to be done to explore the phenomenological consequences of this transition, but our findings suggest that applying the $N_c \rightarrow \infty$ limit to dense baryonic systems should be done with caution, since a discontinuity might be present between $N_c=3$ and $N_c \rightarrow \infty$.

We thank J. Noronha, F. Giacosa, M. Panero, B. Betz, Larry McLerran and Toru Kojo for discussions.
G.~T.~acknowledges the financial support received from the Helmholtz International
Centre for FAIR within the framework of the LOEWE program
(Landesoffensive zur Entwicklung Wissenschaftlich-\"Okonomischer
Exzellenz) launched by the State of Hesse.


\begin{thebibliography}{100}

\bibitem{perc1}
	G.~Baym,
	Physica A {\bf 96}, 131 (1979).

\bibitem{perc2}
	T.~Celik, F.~Karsch, H.~Satz,
	Phys.~Lett.~B {\bf 97}, 128 (1980).

\bibitem{perc3}
  H.~Satz,
  arXiv:hep-ph/0212046.

\bibitem{perc4}
  T.~Tarnowsky, R.~Scharenberg, B.~Srivastava,
  Int.\ J.\ Mod.\ Phys.\  E {\bf 16}, 1859 (2007).


\bibitem{perc5}
  K.~Werner,
  Nucl.\ Phys.\  A {\bf 572}, 141 (1994).


\bibitem{perc6}
  N.~Armesto, M.~A.~Braun, E.~G.~Ferreiro, C.~Pajares,
  Phys.\ Rev.\ Lett.\  {\bf 77}, 3736 (1996)

\bibitem{perc7}
	F.~Gliozzi, S.~Lottini, M.~Panero, A.~Rago,
	Nucl.~Phys.~B {\bf 719}, 255 (2005).

\bibitem{florkowski}
	W.~Florkowski,
	Phenomenology of Ultra-Relativistic Heavy-Ion Collisions,
	{\it World Scientific Publishing Company, Singapore (2010) 436pp.}

\bibitem{low1}
	M.~Posiadala [NA61 Collaboration],
	arXiv:0901.3332. 

\bibitem{low2}
	A.~N.~Sissakian, A.~S.~Sorin [NICA Collaboration],
	J.~Phys.~G {\bf 36}, 064069 (2009).

\bibitem{low3}
	G.~Odyniec,
	Acta Phys.~Polon.~B {\bf 40}, 1237 (2009).

\bibitem{low4}
	P.~Staszel [CBM Collaboration],
	Acta Phys.~Polon.~B {\bf 41}, 341 (2010).

\bibitem{critical}
	M.~A.~Stephanov, K.~Rajagopal, E.~V.~Shuryak,
	Phys.~Rev.~Lett.~{\bf 81}, 4816 (1998).

\bibitem{spino}
	I.~N.~Mishustin,
	arXiv:hep-ph/0512366.


\bibitem{hansen}
	P.~Costa, M.~C.~Ruivo, C.~A.~de Sousa, H.~Hansen,
	Symmetry {\bf 2}, 1338 (2010).

\bibitem{ratti}
	S.~Roessner, T.~Hell, C.~Ratti, W.~Weise,
	Nucl.~Phys.~A {\bf 814}, 118 (2008).

\bibitem{dirkq}
	D.~Rischke,
	PoS Confinement8 (2008),
	pp.137.

\bibitem{sasaki}
	L.~McLerran, K.~Redlich, C.~Sasaki,
	Nucl.~Phys.~A {\bf 824}, 86 (2009).

\bibitem{kojo}
  T.~Kojo,
  arXiv:1106.2187 [hep-ph].

\bibitem{chiralspiral}
	T.~Kojo, R.~D.~Pisarski, A.~M.~Tsvelik,
	Phys.~Rev.~D {\bf 82}, 074015 (2010);
	T.~Kojo, Y.~Hidaka, L.~McLerran, R.~D.~Pisarski,
	Nucl.~Phys.~A {\bf 843}, 37 (2010).

\bibitem{buballa}
	S.~Carignano, D.~Nickel, M.~Buballa,
	Phys.~Rev.~D {\bf 82}, 054009 (2010).

\bibitem{quarkyonic}
	L.~McLerran, R.~D.~Pisarski,
	Nucl.~Phys.~A {\bf 796}, 83 (2007).
 
\bibitem{latmu1}
	Z.~Fodor, S.~D.~Katz,
	Phys.~Lett.~B {\bf 534}, 87 (2002).

\bibitem{latmu2}
	C.~R.~Allton, M.~Doering, S.~Ejiri, S.~J.~Hands, O.~Kaczmarek, F.~Karsch, E.~Laermann, K.~Redlich,
	Phys.~Rev.~D {\bf 71}, 054508 (2005).

\bibitem{latmu3}
	P.~de Forcrand, O.~Philipsen,
	Nucl.~Phys.~B {\bf 642}, 290 (2002).

\bibitem{thooft}
	G.~'t Hooft,
	Nucl.~Phys.~B {\bf 72}, 461 (1974).

\bibitem{witten}
	E.~Witten,
	Nucl.~Phys.~B {\bf 160}, 57 (1979).


\bibitem{maldacena}
	J.~M.~Maldacena,
  arXiv:hep-th/0309246.

\bibitem{sugimoto}
	T.~Sakai, S.~Sugimoto,
	Prog.~Theor.~Phys.~{\bf 113}, 843 (2005).

\bibitem{lippert}
	O.~Bergman, G.~Lifschytz, M.~Lippert,
	JHEP {\bf 0711}, 056 (2007).

\bibitem{rozali}
	M.~Rozali, H.~H.~Shieh, M.~Van Raamsdonk, J.~Wu,
	JHEP {\bf 0801}, 053 (2008).

\bibitem{crystal1}
	I.~Zahed, G.~E.~Brown,
	Phys.~Rept.~{\bf 142}, 1 (1986).


\bibitem{crystal3}
	I.~R.~Klebanov,
	Nucl.~Phys.~B {\bf 262}, 133 (1985).


\bibitem{percthresholds}
	C.~D.~Lorenz, R.~May, R.~M.~Ziff,
	J.~Stat.~Phys.~{\bf 98} 3--4, 961 (2000);
	C.~D.~Lorenz, R.~M.~Ziff,
	Phys.~Rev.~E {\bf 57}, 230 (1998).

\bibitem{ncgiac}
	L.~Bonanno, F.~Giacosa,
	arXiv:1102.3367 [hep-ph].

\bibitem{ncus}
	G.~Torrieri, I.~Mishustin,
	Phys.~Rev.~C {\bf 82}, 055202 (2010).

\bibitem{lat1}
	B.~Bringoltz, M.~Teper,
	Phys.~Lett.~B {\bf 628}, 113 (2005).

	\bibitem{lat2}
	M.~Panero,
	Phys.~Rev.~Lett.~{\bf 103}, 232001 (2009).

\bibitem{fromm}
	P.~de Forcrand, M.~Fromm,
	Phys.~Rev.~Lett.~{\bf 104}, 112005 (2010).

\bibitem{stringperc}
	G.~R.~Harris,
	Nucl.~Phys.~B {\bf 418}, 278 (1994)


\bibitem{nicolini}
  P.~Nicolini and G.~Torrieri,
  arXiv:1105.0188 [gr-qc].

\end{thebibliography}
\end{document}